\documentclass{article}
\usepackage[utf8]{inputenc}
\usepackage{amsmath}
\usepackage{amssymb}
\usepackage{graphicx}
\usepackage{lscape}
\usepackage{algorithm}
\usepackage{algorithmic}

\title{A Fully Bayesian, Logistic Regression Tracking Algorithm for Mitigating Disparate Misclassification}
\author{M.B. Short, G.O. Mohler}
\date{\today}

\begin{document}

\maketitle

\begin{abstract}
We develop a fully Bayesian, logistic tracking algorithm with the purpose of providing classification results that are unbiased when applied uniformly to individuals with differing sensitive variable values.  Here, we consider bias in the form of differences in false prediction rates between the different sensitive variable groups.  Given that the method is fully Bayesian, it is well suited for situations where group parameters or logistic regression coefficients are dynamic quantities.  We illustrate our method, in comparison to others, on both simulated datasets and the well-known ProPublica COMPAS dataset.
\end{abstract}

\section{Introduction}

Algorithmic scoring is employed in a variety of decision making situations including parole and bail \cite{dressel2018accuracy,berk2017impact}, loan approval \cite{mothilal2020explaining}, and credit scoring \cite{wang2011comparative}.  In the case of bail decisions and the well-known COMPAS algorithm, false positive rates are much higher for African-American defendants compared to Caucasian \cite{dressel2018accuracy}.  Recently a number of approaches have been introduced to improve fairness of machine learning algorithms. In \cite{hardt2016equality}, disparate group thresholds on logistic regression predictions are used to improve fairness post model training.  In \cite{zafar2017fairness}, a penalized loss is used in training to match false positive and negative rates across a sensitive variable and in \cite{lum2016statistical} the authors transform input features to achieve independence of predictions from a sensitive variable.  In all of these studies offline training is used and predictions are evaluated in batch on a test set.  Furthermore, these methods do not usually quantify uncertainty in estimates of fairness or disparate impact.

In this paper we introduce a method for mitigating disparate impact in logistic regression where the distribution of covariates may be changing over time and need to be tracked.  Under this scenario, a fairness-aware algorithm trained offline may deviate from fair predictions over time due to the changing distribution of the data.  We therefore propose a Bayesian tracking method that can estimate changing covariates in real-time and dynamically modify decisions to mitigate disparate impact.  The method has the additional benefit of allowing for uncertainty quantification in fairness-aware logistic regression.  To date Bayesian approaches to fairness have been limited to offline studies \cite{dimitrakakis2019bayesian,simoiu2017problem} and to our knowledge this is the first to consider the Bayesian fairness tracking problem.

It is known \cite{zhang2017causal} that classification algorithms can lead to outcomes that display certain kinds of bias when the distribution of feature vectors varies from one sensitive group to the next, even if the algorithm does not explicitly include knowledge of the sensitive variables when making its classifications.  One could attempt to counteract this by having the algorithm explicitly take into account the sensitive variable of the individual in question when making a prediction, for the purposes of removing the implicit bias.  However, this is often deemed as undesirable, and in some real-use cases, may be illegal.  Instead, we seek a tracking algorithm that does not explicitly take into account an individual's sensitive variable value in order to make the prediction, but instead applies the exact same method uniformly to all individuals, while nonetheless attempting to guarantee similar statistical results across groups.

The outline of the paper is as follows.  In Section 2 we present the formulation of the problem that we study and the details of our Bayesian logistic tracking algorithm.  In Section 3 we demonstrate the effectiveness of the approach on synthetic data when ground truth is known and in Section 4 we illustrate the application of the methodology to the well-known ProPublica COMPAS dataset.  We discuss our results and directions for future research in Section 5.

\section{Methodology}
Our algorithm shall accept as input streaming, $N$ dimensional feature data $\mathbf{x}_i$, where subscript $i$ denotes the time at which this specific data point arrives for processing; we assume that no two individual's data arrives simultaneously, so $i$ also implicitly references individuals as well.  It is not strictly necessary that the data actually be generated at different, well-ordered points in time.  However, as a tracking algorithm, one of the strengths of this method is that it is built to handle data that is dynamically evolving in some way, so we cast the problem in this light.  As one of our main concerns here is providing an algorithm that is unbiased when applied to data from individuals with different sensitive parameter values, we stipulate that each feature vector $\textbf{x}_i$ is accompanied by a categorical value $z_i$ that indicates the value of a sensitive variable (sex, race, age, etc.) for the individual $i$.  The algorithm's goal is to produce binary classifications $\hat{y}_i\in\{0,1\}$ for each individual; depending on the domain in question, this classification could correspond to a belief that the individual will or will not default on a loan, commit a crime in the near future, or soon become homeless, among many other possibilities.  We differentiate here between the predicted classifications $\hat{y}_i$ and the true classifications $y_i$.  Note that in some domains, the true classifications may not always be available, or even in some sense exist, or may only become available after some time has passed after the predicted classification is made.  For the purposes of this study, we simply assume that $y_i$ exists and is known immediately after the predicted value $\hat{y}_i$ is generated.

\subsection{Bayesian logistic tracker}
To perform the classification, we employ a Bayesian logistic tracker.  The logistic model assumes that the true classifications $y_i$ are Bernoulli random variables with probability
\begin{equation} \label{eq:logistic}
p(\mathbf{x}_i|\mathbf{\theta}_i)=\frac{e^{\mathbf{\theta}_i^T\mathbf{x}_i}}{1+e^{\mathbf{\theta}_i^T\mathbf{x}_i}}~,
\end{equation}
where $\mathbf{\theta}_i$ is an $N$ dimensional vector of feature weights at time $i$.  Then our algorithm will attempt to recursively generate an estimate for $\mathbf{\theta}_{i}$, in the form of a probability distribution, given a sequence of observations $\{\mathbf{x}_j\}$ and $\{y_j\}$ for $j\leq i$.

Noting that the model \eqref{eq:logistic} is nonlinear, two prominent possibilities to perform the tracking are the Unscented Kalman filter and the Extended Kalman filter.  For our purposes, we employ an Extended Kalman filter, though we make no claim of its superiority over the Unscented Kalman filter, other than it's relative speed in our particular case.  To develop our Extended Kalman filter, we first note that the probability of observing a specific $y_i$ is given by
\[
P(y_i|\mathbf{\theta}_i,\mathbf{x}_i)=p(\mathbf{x}_i|\mathbf{\theta}_i)^{y_i}\left(1-p(\mathbf{x}_i|\mathbf{\theta}_i)\right)^{1-y_i}~.
\]
Let us now assume that the prior belief over $\mathbf{\theta}_i$ before incorporating observation $y_i$ is a multivariate Normal with mean $\overline{\mathbf{\theta}}_{i|i-1}$ and covariance matrix $C_{i|i-1}$, denoted $\mathcal{N}(\mathbf{\theta}_i|\overline{\mathbf{\theta}}_{i|i-1},C_{i|i-1})$.  We further insist that the posterior belief over $\mathbf{\theta}_i$ after incorporating observation $y_i$ is a multivariate Normal with mean $\overline{\mathbf{\theta}}_{i}$ and covariance matrix $C_{i}$, $\mathcal{N}(\mathbf{\theta}_i|\overline{\mathbf{\theta}}_{i},C_{i})$
.  Then Bayes' rule, with logarithms applied to both sides of the equation, gives
\begin{multline} \label{eq:logbayes}
-\frac{1}{2}\left(\mathbf{\theta}_i-\overline{\mathbf{\theta}}_i\right)^TC_i^{-1}\left(\mathbf{\theta}_i-\overline{\mathbf{\theta}}_i\right)=y_i\ln\left[p(\mathbf{x}_i|\mathbf{\theta}_i)\right]+\\
(1-y_i)\ln\left[1-p(\mathbf{x}_i|\mathbf{\theta}_i)\right]-\frac{1}{2}\left(\mathbf{\theta}_i-\overline{\mathbf{\theta}}_{i|i-1}\right)^TC_{i|i-1}^{-1}\left(\mathbf{\theta}_i-\overline{\mathbf{\theta}}_{i|i-1}\right)+D~,
\end{multline}
where $D$ is a constant unrelated to $\mathbf{\theta}_i$. We now Taylor expand the logarithmic terms on the right hand side of \eqref{eq:logbayes} up to second order around the point $\mathbf{\theta}_i=\overline{\mathbf{\theta}}_{i|i-1}$ and, after some algebraic manipulation, obtain our iterative update equations by matching the linear and quadratic terms in $\mathbf{\theta}_i$ on both sides of the equation, finding
\begin{align}
C_i^{-1}&=C_{i|i-1}^{-1}+\mathbf{h}_i\mathbf{h}_i^T~,\label{eq:covupdate1}\\
\overline{\mathbf{\theta}}_i&=\overline{\mathbf{\theta}}_{i|i-1}-C_i\mathbf{f}_i \label{eq:meanupdate}
\end{align}
where
\begin{align}
\mathbf{f}_i&=(-1)^{y_i}\mathbf{x}_ip\left((-1)^{y_i}\mathbf{x}_i|\overline{\mathbf{\theta}}_{i|i-1}\right)~,\\
\mathbf{h}_i&=\mathbf{x}_i p\left((-1)^{1-y_i}\mathbf{x}_i\right|\overline{\mathbf{\theta}}_{i|i-1})\exp\left((-1)^{y_i}\overline{\mathbf{\theta}}_{i|i-1}^T\mathbf{x}_i/2\right)~.
\end{align}
Because of the special form of the matrix used to update $C_i^{-1}$ in \eqref{eq:covupdate1}, the equation \eqref{eq:covupdate1} can be computed very efficiently without the need for any matrix inversion by using the rank-one update rule
\begin{equation} \label{eq:covupdate2}
C_i=C_{i|i-1}-\frac{C_{i|i-1}\mathbf{h}_i\mathbf{h}_i^TC_{i|i-1}}{1+\mathbf{h}_i^TC_{i|i-1}\mathbf{h}_i}~.
\end{equation}

The tracking algorithm is completed by providing a model for the dynamics of $\mathbf{\theta}_i$, allowing one to find the prior parameters $\overline{\mathbf{\theta}}_{i|i-1}$ and $C_{i|i-1}$ from the posterior parameters $\overline{\mathbf{\theta}}_{i-1}$ and $C_{i-1}$ from the previous observation.  For our purposes, and lacking any more informed model, we simply make the common choice that $\mathbf{\theta}_i$ is undergoing a simple random walk  $\mathbf{\theta}_i=\mathbf{\theta}_{i-1}+\mathcal{N}(\mathbf{0},Q)$; this choice maintains normality of the probability distribution.  Then we simply have $\overline{\mathbf{\theta}}_{i|i-1}=\overline{\mathbf{\theta}}_{i-1}$ and $C_{i|i-1}=C_{i-1}+Q$.  The value of covariance matrix $Q$ affects how well the algorithm is able to track changes over time, with too high a value causing the tracked value to fluctuate too rapidly and too low a value causing the system to adjust too slowly to changes in the tracked variable.

\subsection{Bias estimation and elimination}
Given our streaming data and the output of our Bayesian tracker \eqref{eq:meanupdate}-\eqref{eq:covupdate2}, one can construct predictions $\hat{y}_i$ for the classifications $y_i$ by first computing the expected probability for an individual with feature vector $\mathbf{x}_i$,
\begin{equation}
    \overline{p}(\mathbf{x}_i|\overline{\mathbf{\theta}}_{i|i-1},C_{i|i-1})=\int_{\mathbb{R}^N}p(\mathbf{x}_i|\mathbf{\theta}_i)\mathcal{N}(\mathbf{\theta}_i|\overline{\mathbf{\theta}}_{i|i-1},C_{i|i-1}) d\mathbf{\theta}_i~,
\end{equation}
and then thresholding this probability by a value $\tau$ such that $\hat{y}_i=1$ if $\overline{p}>\tau$ and $\hat{y}_i=0$ if $\overline{p}<\tau$; generally, and in the remainder of this work, $\tau=0.5$.  A simpler version of the integral above can be found by first noting that $p(\mathbf{x}_i|\mathbf{\theta}_i)$ depends only on the argument $q_i=\mathbf{\theta}_i^T\mathbf{x}_i$ and employing the well known property that if $\mathbf{\theta}_i\sim \mathcal{N}(\overline{\mathbf{\theta}}_{i|i-1},C_{i|i-1})$, then $\mathbf{\theta}_i^T\mathbf{x}_i\sim \mathcal{N}(\overline{\mathbf{\theta}}_{i|i-1}^T\mathbf{x}_i,\mathbf{x}_i^TC_{i|i-1}\mathbf{x}_i)$.  Then we have
\begin{equation}
    \overline{p}(\mathbf{x}_i|\overline{\mathbf{\theta}}_{i|i-1},C_{i|i-1})=\int_{-\infty}^\infty \frac{e^{q_i}}{1+e^{q_i}}\mathcal{N}(q_i|\overline{\mathbf{\theta}}_{i|i-1}^T\mathbf{x}_i,\mathbf{x}_i^TC_{i|i-1}\mathbf{x}_i) dq_i~.
\end{equation}
The above integral has several known approximations that can be used to simplify computation, and we have chosen to use the approximation from \cite{logistic-approx}.

However, as previously mentioned, under many circumstances these predictions will show bias in terms of false prediction rates when comparing between those predictions made for individuals of differing sensitive variable values.  For example, consider the case in which the sensitive variable value $z_i$ may only take one of two possible values, which we will simply choose to be 0 and 1 for convenience.  Let us assume that the feature vectors $\mathbf{x}_i$ for those individuals with sensitive variable value 0 are well described by a probability density $\mathcal{D}_0(\mathbf{x}_i)$ and similarly $\mathcal{D}_1(\mathbf{x}_i)$ for individuals with sensitive variable value 1.  Then if we were to employ posteriors $\overline{\mathbf{\theta}}_{i}$ and $C_{i}$ to make hypothetical predictions for more individuals at time $i$, the expected instantaneous false negative rate and false positive rate for sensitive variable value $z$, $\textrm{FNR}_i(z)$ and $\textrm{FPR}_i(z)$ respectively, would be
\begin{align}
    \textrm{FNR}_i(z)&=\frac{\int_{\mathbb{R}^N}\mathbf{1}_{\overline{p}(\mathbf{x}_i|\overline{\mathbf{\theta}}_{i},C_{i})<\tau}\overline{p}(\mathbf{x}_i|\overline{\mathbf{\theta}}_{i},C_{i})\mathcal{D}_z(\mathbf{x}_i)d\mathbf{x}_i }{\int_{\mathbb{R}^N}\overline{p}(\mathbf{x}_i|\overline{\mathbf{\theta}}_{i},C_{i})\mathcal{D}_z(\mathbf{x}_i)d\mathbf{x}_i}~,\\
    \textrm{FPR}_i(z)&=\frac{\int_{\mathbb{R}^N}\mathbf{1}_{\overline{p}(\mathbf{x}_i|\overline{\mathbf{\theta}}_{i},C_{i})>\tau}[1-\overline{p}(\mathbf{x}_i|\overline{\mathbf{\theta}}_{i},C_{i})]\mathcal{D}_z(\mathbf{x}_i)d\mathbf{x}_i }{\int_{\mathbb{R}^N}[1-\overline{p}(\mathbf{x}_i|\overline{\mathbf{\theta}}_{i},C_{i})]\mathcal{D}_z(\mathbf{x}_i)d\mathbf{x}_i}~,
\end{align}
where $\mathbf{1}$ is an indicator function.  Similarly, the expected instantaneous accuracy would be given by 
\begin{multline}
    \textrm{ACC}_i(z)=\int_{\mathbb{R}^N}\mathbf{1}_{\overline{p}(\mathbf{x}_i|\overline{\mathbf{\theta}}_{i},C_{i})>\tau}\overline{p}(\mathbf{x}_i|\overline{\mathbf{\theta}}_{i},C_{i})\mathcal{D}_z(\mathbf{x}_i)d\mathbf{x}_i~+\\
    \int_{\mathbb{R}^N}\mathbf{1}_{\overline{p}(\mathbf{x}_i|\overline{\mathbf{\theta}}_{i},C_{i})<\tau}\left[1-\overline{p}(\mathbf{x}_i|\overline{\mathbf{\theta}}_{i},C_{i})\right]\mathcal{D}_z(\mathbf{x}_i)d\mathbf{x}_i~.\label{eq:acct}
\end{multline}
It is important to note that discrepancies between the expected false prediction rates for the two sensitive variable values can arise even if the posterior distribution parameters $\overline{\mathbf{\theta}}_{i}$ and $C_{i}$ are absolutely correct; that is, even if they themselves are not biased due to flawed data being used to estimate them.  This is simply due to the fact that $\mathcal{D}_0(\mathbf{x}_i)$ and $\mathcal{D}_1(\mathbf{x}_i)$ may differ \cite{zhang2017causal}.

If we assume that $\overline{\mathbf{\theta}}_{i}$ and $C_{i}$ are indeed correct, but the false prediction rates between the two groups are unequal, then the only way to possibly equalize the false prediction rates is to change the indicator function term present in the integrals above, which represents the prediction methodology.  
The method we choose to equalize false prediction rates is to base our predictions not off of $\overline{\mathbf{\theta}}_{i}$ and $C_{i}$, but off of some other parameters $\overline{\mathbf{\Theta}}_{i}$ and $\mathbb{C}_{i}$ instead.  Then our false prediction rates and accuracy are 
\begin{align}
    \hat{\textrm{FNR}}_i(z)&=\frac{\int_{\mathbb{R}^N}\mathbf{1}_{\overline{p}(\mathbf{x}_i|\overline{\mathbf{\Theta}}_{i},\mathbb{C}_{i})<\tau}\overline{p}(\mathbf{x}_i|\overline{\mathbf{\theta}}_{i},C_{i})\mathcal{D}_z(\mathbf{x}_i)d\mathbf{x}_i }{\int_{\mathbb{R}^N}\overline{p}(\mathbf{x}_i|\overline{\mathbf{\theta}}_{i},C_{i})\mathcal{D}_z(\mathbf{x}_i)d\mathbf{x}_i}~,\label{eq:fnr}\\
    \hat{\textrm{FPR}}_i(z)&=\frac{\int_{\mathbb{R}^N}\mathbf{1}_{\overline{p}(\mathbf{x}_i|\overline{\mathbf{\Theta}}_{i},\mathbb{C}_{i})>\tau}[1-\overline{p}(\mathbf{x}_i|\overline{\mathbf{\theta}}_{i},C_{i})]\mathcal{D}_z(\mathbf{x}_i)d\mathbf{x}_i }{\int_{\mathbb{R}^N}[1-\overline{p}(\mathbf{x}_i|\overline{\mathbf{\theta}}_{i},C_{i})]\mathcal{D}_z(\mathbf{x}_i)d\mathbf{x}_i}~.\label{eq:fpr}
\end{align}
\begin{multline}
    \hat{\textrm{ACC}}_i(z)=\int_{\mathbb{R}^N}\mathbf{1}_{\overline{p}(\mathbf{x}_i|\overline{\mathbf{\Theta}}_{i},\mathbb{C}_{i})>\tau}\overline{p}(\mathbf{x}_i|\overline{\mathbf{\theta}}_{i},C_{i})\mathcal{D}_z(\mathbf{x}_i)d\mathbf{x}_i~+\\
    \int_{\mathbb{R}^N}\mathbf{1}_{\overline{p}(\mathbf{x}_i|\overline{\mathbf{\Theta}}_{i},\mathbb{C}_{i})<\tau}\left[1-\overline{p}(\mathbf{x}_i|\overline{\mathbf{\theta}}_{i},C_{i})\right]\mathcal{D}_z(\mathbf{x}_i)d\mathbf{x}_i~.\label{eq:accf}
\end{multline}
The goal then is to generate posterior parameters $\overline{\mathbf{\Theta}}_{i}$ and $\mathbb{C}_{i}$ that reduce or eliminate bias, while still retaining some level of accuracy.

To detail how we accomplish this, we begin first by defining our precise metric for measuring expected bias at time $i$
\begin{equation}
    \Delta_i=\sqrt{[\hat{\textrm{FPR}}_i(1)-\hat{\textrm{FPR}}_i(0)]^2+[\hat{\textrm{FNR}}_i(1)-\hat{\textrm{FNR}}_i(0)]^2}~,\label{metric}
\end{equation}
where our goal will be to make $\Delta_i<\epsilon$ for some chosen small $\epsilon$ value.  We note that other fairness metrics may be used in place of Equation \ref{metric}.  See \cite{mehrabi2019survey} for a review of fair machine learning, including a summary of the different metrics used in practice, and \cite{corbett2018measure} for a discussion of the tradeoffs.

Suppose, then, that we possess some prior values for $\overline{\mathbf{\Theta}}_{i|i-1}$ and $\mathbb{C}_{i|i-1}$.  Given data point $\mathbf{x}_i$, $y_i$, we could use our Bayesian tracker to update them, via
\begin{align}
\mathbb{C}_i&=\mathbb{C}_{i|i-1}-\frac{\mathbb{C}_{i|i-1}\mathbf{h}_i\mathbf{h}_i^T\mathbb{C}_{i|i-1}}{1+\mathbf{h}_i^T\mathbb{C}_{i|i-1}\mathbf{h}_i}~,\label{eq:covfair}\\
\overline{\mathbf{\Theta}}_i&=\overline{\mathbf{\Theta}}_{i|i-1}-\mathbb{C}_i\mathbf{f}_i
\end{align}
where
\begin{align}
\mathbf{f}_i&=(-1)^{y_i}\mathbf{x}_ip\left((-1)^{y_i}\mathbf{x}_i|\overline{\mathbf{\Theta}}_{i|i-1}\right)~,\\
\mathbf{h}_i&=\mathbf{x}_i p\left((-1)^{1-y_i}\mathbf{x}_i\right|\overline{\mathbf{\Theta}}_{i|i-1})\exp\left((-1)^{y_i}\overline{\mathbf{\Theta}}_{i|i-1}^T\mathbf{x}_i/2\right)~.\label{eq:hfair}
\end{align}
This posterior distribution may already meet our bias metric, in which case we could use it to construct the prior for the next observation $i+1$ and proceed as usual with our tracker.  On the other hand, these new values could lead to a posterior distribution that violates the bias metric.  However, even in this case it is very likely that some regions of this posterior distribution would not violate the metric.  Then, what we seek is to alter our posterior distribution by retaining only that region over which the bias metric would be satisfied.  In practice, we achieve this via Monte Carlo sampling.  Specifically, we sample $M_\Theta$ potential predictor coefficient vectors $\tilde{\mathbf{\Theta}}_{ij}$ from the posterior $\mathcal{N}(\overline{\mathbf{\Theta}}_i,\mathbb{C}_i)$ and for each of these we calculate $\hat{\textrm{FNR}}_{ij}(z)$ and $\hat{\textrm{FPR}}_{ij}(z)$ using \eqref{eq:fpr}-\eqref{eq:fnr} but with the indicator function replaced with $\mathbf{1}_{p(\mathbf{x}_i|\tilde{\mathbf{\Theta}}_{ij})}$ -- that is, we use the sampled predictor coefficient vectors to make the hypothetical predictions.  We can then calculate the bias $\Delta_{ij}$ for each sampled vector, retaining those samples whose $\Delta_{ij}<\epsilon$ and rejecting those whose $\Delta_{ij}>\epsilon$.  However, since some such low-bias coefficient vectors may also lead to very low predictive accuracy, we further restrict our samples to those whose predictive accuracy from \eqref{eq:accf}, with indicator functions replaced with $\mathbf{1}_{p(\mathbf{x}_i|\tilde{\mathbf{\Theta}}_{ij})}$, lies above some threshold in relation to the predictive accuracy of \eqref{eq:acct}.  Specifically, we require that
\begin{equation}
    \min_z\left[ \hat{\textrm{ACC}}_{ij}(z)/\textrm{ACC}_{i}(z)\right] > \alpha~,\label{eq:accmet}
\end{equation}
where $0<\alpha<1$.  After processing all $M_\Theta$ samples in this way, the mean $\overline{\mathbf{\Theta}}_i^\epsilon$ and covariance matrix $\mathbb{C}_i^\epsilon$ for that subset meeting our bias and accuracy criteria is computed, and we simply use those to construct our prior for the next step in the tracker via $\overline{\mathbf{\Theta}}_{i+i|i}= \overline{\mathbf{\Theta}}_i^\epsilon$ and $\mathbb{C}_{i+1|i}=\mathbb{C}_i^\epsilon+Q$.

We see, then, that our method involves tracking two (potentially) different coefficient vector distributions using \eqref{eq:covupdate1}-\eqref{eq:covupdate2}: the ``true'' distribution $\mathcal{N}(\mathbf{\theta}_i|\overline{\mathbf{\theta}}_{i},C_{i})$ that is never used to make predictions $\hat{y}$ but is used to predict the current expected bias level and accuracy, and the ``unbiased'' distribution $\mathcal{N}(\mathbf{\theta}_i|\overline{\mathbf{\Theta}}_{i}^\epsilon,\mathbb{C}_{i}^\epsilon)$ that is used to make predictions $\hat{y}$ and is forced to meet our bias threshold $\epsilon$ after every datapoint $y_i$ is assimilated via the sampling method described above, while retaining at least some relative measure of accuracy.

Of course, the sampling method for our unbiased distribution involves evaluating the integrals in \eqref{eq:fpr}-\eqref{eq:accf}, which is easier said than done.  There are at least two difficulties: the integral may be over a high dimensional space and the integrals require the knowledge of $\mathcal{D}_z(\mathbf{x}_i)$.  We deal with the first problem by performing the integral via Monte Carlo methods.  That is, we generate $M_x$ feature vectors $\tilde{\mathbf{x}}_k$ from the relevant $\mathcal{D}$, then for each sample we determine its true probability $\overline{p}(\tilde{\mathbf{x}}_k|\overline{\mathbf{\theta}}_i,C_i)$ and its predicted probability using $p(\tilde{\mathbf{x}}_k|\tilde{\mathbf{\Theta}}_{ij})$.  Then the integrals are approximated as
\begin{align}
    \hat{\textrm{FNR}}_{ij}(z)&=\frac{\sum_{k}\mathbf{1}_{p(\tilde{\mathbf{x}}_k|\tilde{\mathbf{\Theta}}_{ij})<\tau}\overline{p}(\tilde{\mathbf{x}}_k|\overline{\mathbf{\theta}}_{i},C_{i})}{\sum_{k}\overline{p}(\tilde{\mathbf{x}}_k|\overline{\mathbf{\theta}}_{i},C_{i})}~,\label{eq:fnr2}\\
    \hat{\textrm{FPR}}_{ij}(z)&=\frac{\sum_{k}\mathbf{1}_{p(\tilde{\mathbf{x}}_k|\tilde{\mathbf{\Theta}}_{ij})>\tau}\left[1-\overline{p}(\tilde{\mathbf{x}}_k|\overline{\mathbf{\theta}}_{i},C_{i})\right]}{\sum_{k}\left[1-\overline{p}(\tilde{\mathbf{x}}_k|\overline{\mathbf{\theta}}_{i},C_{i})\right]}~,\label{eq:fpr2}
\end{align}
\begin{multline}
    \hat{\textrm{ACC}}_{ij}(z)=\frac{1}{M_x}\sum_{k}\mathbf{1}_{p(\tilde{\mathbf{x}}_k|\tilde{\mathbf{\Theta}}_{ij})>\tau}\overline{p}(\tilde{\mathbf{x}}_k|\overline{\mathbf{\theta}}_{i},C_{i})~+\\
    \frac{1}{M_x}\sum_{k}\mathbf{1}_{p(\tilde{\mathbf{x}}_k|\tilde{\mathbf{\Theta}}_{ij})<\tau}\left[1-\overline{p}(\tilde{\mathbf{x}}_k|\overline{\mathbf{\theta}}_{i},C_{i})\right]~;\label{eq:acc2}
\end{multline}
the integral within \eqref{eq:acct} is evaluated similarly.

\subsection{Bayesian feature tracker}
The only remaining portion of our algorithm to describe is our method for estimating the distributions $\mathcal{D}_z$ for the feature vectors $\mathbf{x}$ of individuals with sensitive variable value $z$.  First, we note that the $N$ dimensional feature vectors will generally all have an entry of 1 as their final component, essentially allowing for a constant probability offset for all individuals in the dataset if necessary, as is standard in logistic regression.  This means that the distributions $\mathcal{D}_z$ are really $N-1$ dimensional.  To track them, we use a standard approach, assuming that the first $N-1$ entries of each $\mathbf{x}$ for a given sensitive variable value $z$ are drawn from a normal distribution $\mathcal{N}(\mathbf{x}|\mathbf{\mu}_{z,i},\Sigma_{z,i})$.  The parameters $\mathbf{\mu}_{z,i}$ and $\Sigma_{z,i}$ are themselves unknown, but we select their prior to be a normal-inverse-Wishart distribution with hyperparameters $\mathbf{m}_{z,i|i-1}$, $\lambda_{z,i|i-1}$, $\Phi_{z,i|i-1}$, and $\nu_{z,i|i-1}$; note that it is required that $\lambda_{z,i|i-1}>0$, $\nu_{z,i|i-1}>N-2$, and $\Phi_{z,i|i-1}$ be a positive definite $N-1\times N-1$ matrix. This choice is the conjugate prior of the multinormal distribution assumed for the observations $\mathbf{x}$, so that a relatively simple Bayesian update rule for the posterior hyperparameters is known:
\begin{align}
\lambda_{z,i}&=\lambda_{z,i|i-1}+1\label{eq:lambda}\\
\mathbf{m}_{z,i}&=\frac{\lambda_{z,i|i-1}\mathbf{m}_{z,i|i-1}+\mathbf{x}_i}{\lambda_{z,i}}\\
\Phi_{z,i}&=\Phi_{z,i|i-1}+\frac{\lambda_{z,i|i-1}}{\lambda_{z,i}}(\mathbf{x}_i-\mathbf{m}_{z,i|i-1})(\mathbf{x}_i-\mathbf{m}_{z,i|i-1})^T\\
\nu_{z,i}&=\nu_{z,i|i-1}+1 \label{eq:nu}
\end{align}

Given these posterior hyperparameters, the posterior predictive distribution $\hat{\mathcal{D}}_z$ for $\mathbf{x}$ for sensitive variable value $z$ is multivariate $t$:
\begin{equation} \label{eq:postpred}
    \hat{\mathcal{D}}_z(\mathbf{x}_i)=t_{\nu_{z,i}-N+2}\left(\mathbf{x}_i|\mathbf{m}_{z,i},\frac{\lambda_{z,i}+1}{\lambda_{z,i}(\nu_{z,i}-N+2)}\Phi_{z,i}\right)~.
\end{equation}
It is this posterior predictive distribution that is used to generate the samples $\tilde{\mathbf{x}}_k$ used in \eqref{eq:fpr2}-\eqref{eq:acc2}.  For the multivariate $t$ to have a finite mean and variance, we need $\nu_{z,i}>N$, which is more restrictive than the requirement above.

We complete the specification of the feature vector distribution tracker by providing a means by which the posterior parameters values after step $i$ are used to construct the prior for step $i+1$.
We use $\mathbf{m}_{z,i|i-1}=\mathbf{m}_{z,i-1}$, $\Phi_{z,i|i-1}=\Phi_{z,i-1}$, and $\nu_{z,i|i-1}=\nu_{z,i-1}$.  Since $\lambda$ effectively serves as a factor that weighs how much the prior mean contributes to the posterior mean, and we are interested in scenarios where the mean may be evolving over time, we do not want $\lambda$ to continually increase after each observation, as \eqref{eq:lambda} might indicate.  Hence, we simply set $\lambda_{z,i|i-1}=\beta$, where so long as $\beta\gg 1$, the mean will not fluctuate too rapidly as new observations are made, but tracking an evolving mean will still be possible. 

We summarize the full algorithm in Algorithm \ref{algo}.
\begin{algorithm}
\caption{Bias reducing logistic regression tracking algorithm \label{algo}}
\begin{algorithmic}
\STATE \textbf{Input:} $M$ feature vectors $\mathbf{x}_i$, sensitive variable values $z_i$, and true classifications $y_i$; bias threshold $\epsilon$, accuracy threshold $\alpha$, covariance matrix $Q$, and feature tracker sensitivity $\beta$
\STATE \textbf{Initialize:} priors $\overline{\theta}_{1|0}$, $C_{1|0}$, $\overline{\mathbf{\Theta}}_{1|0}$, $\mathbb{C}_{1|0}$, $\lambda_{z,1|0}$, $\mathbf{m}_{z,1|0}$, $\Phi_{z,1|0}$, $\nu_{z,1|0}$
\FOR{$i=1:M$}
\STATE $Z\leftarrow z_i$
\STATE Find $\lambda_{Z,i}$, $\mathbf{m}_{Z,i}$, $\Phi_{Z,i}$, $\nu_{Z,i}$ using \eqref{eq:lambda}--\eqref{eq:nu}
\FOR{$z\neq Z$}
\STATE Set posteriors $\lambda_{z,i}$, $\mathbf{m}_{z,i}$, $\Phi_{z,i}$, $\nu_{z,i}$ to their current prior values
\ENDFOR
\STATE Find $C_i$ and $\overline{\theta}_i$ using \eqref{eq:meanupdate}--\eqref{eq:covupdate2}
\STATE Find $\mathbb{C}_i$ and $\overline{\mathbf{\Theta}}_i$ using \eqref{eq:covfair}--\eqref{eq:hfair}
\FORALL{$z$}
\STATE Generate set of $M_x$ feature vectors $\{\tilde{\mathbf{x}}\}_z\sim\hat{\mathcal{D}}_z$ from \eqref{eq:postpred}
\STATE Find $\textrm{ACC}_i(z)$ from \eqref{eq:acct} with integrals evaluated as in \eqref{eq:acc2}
\ENDFOR
\STATE $\tilde{\mathbf{\Theta}}_{i0}\leftarrow \overline{\mathbf{\Theta}}_i$
\FORALL{$z$}
\STATE Compute $\hat{\textrm{FNR}}_{i0}(z)$ and $\hat{\textrm{FPR}}_{i0}(z)$ using \eqref{eq:fnr2}--\eqref{eq:fpr2}
\ENDFOR
\STATE Compute $\Delta_{i0}$ using \eqref{metric}
\IF{$\Delta_{i0}<\epsilon$}
\STATE $\overline{\mathbf{\Theta}}_i^\epsilon\leftarrow \overline{\mathbf{\Theta}}_i $, $\mathbb{C}_i^\epsilon\leftarrow\mathbb{C}_i$
\ELSE
\STATE Generate set of $M_\Theta$ coefficient vectors $\{\tilde{\mathbf{\Theta}}_{i}\}\sim\mathcal{N}(\overline{\mathbf{\Theta}}_i,\mathbb{C}_i)$
\FOR{$j=1:M_\Theta$}
\FORALL{$z$}
\STATE Compute $\hat{\textrm{FNR}}_{ij}(z)$, $\hat{\textrm{FPR}}_{ij}(z)$, and $\hat{\textrm{ACC}}_{ij}(z)$ using \eqref{eq:fnr2}--\eqref{eq:acc2}
\ENDFOR
\STATE Compute $\Delta_{ij}$ using \eqref{metric}
\IF{$\Delta_{ij}>\epsilon$ or accuracy goal \eqref{eq:accmet} is violated}
\STATE Remove entry $j$ from set $\{\tilde{\mathbf{\Theta}}_{i}\}$
\ENDIF
\ENDFOR
\STATE $\overline{\mathbf{\Theta}}_i^\epsilon\leftarrow \textrm{mean}\left(\{\tilde{\mathbf{\Theta}}_{i}\}\right)$, $\mathbb{C}_i^\epsilon\leftarrow \textrm{covariance}\left(\{\tilde{\mathbf{\Theta}}_{i}\}\right)$
\ENDIF
\STATE Set priors $\mathbf{m}_{z,i+1|i}$, $\Phi_{z,i+1|i}$, $\nu_{z,i+1|i}$ and $\overline{\theta}_{i+1|i}$ to their current posterior values
\STATE $\overline{\mathbf{\Theta}}_{i+1|i}\leftarrow \overline{\mathbf{\Theta}}_i^\epsilon$, $\mathbb{C}_{i+1|i}\leftarrow\mathbb{C}_i^\epsilon+Q$, $C_{i+1|i}\leftarrow C_i+Q$, $\lambda_{z,i+1|i}\leftarrow \beta$

\ENDFOR
\end{algorithmic}
\end{algorithm}

\section{Results on Synthetic data}
We first illustrate our method on synthetic datasets, both with static and dynamic parameters. 

\subsection{Static parameters}
\begin{figure}
    \centering
    \includegraphics[width=0.49\textwidth]{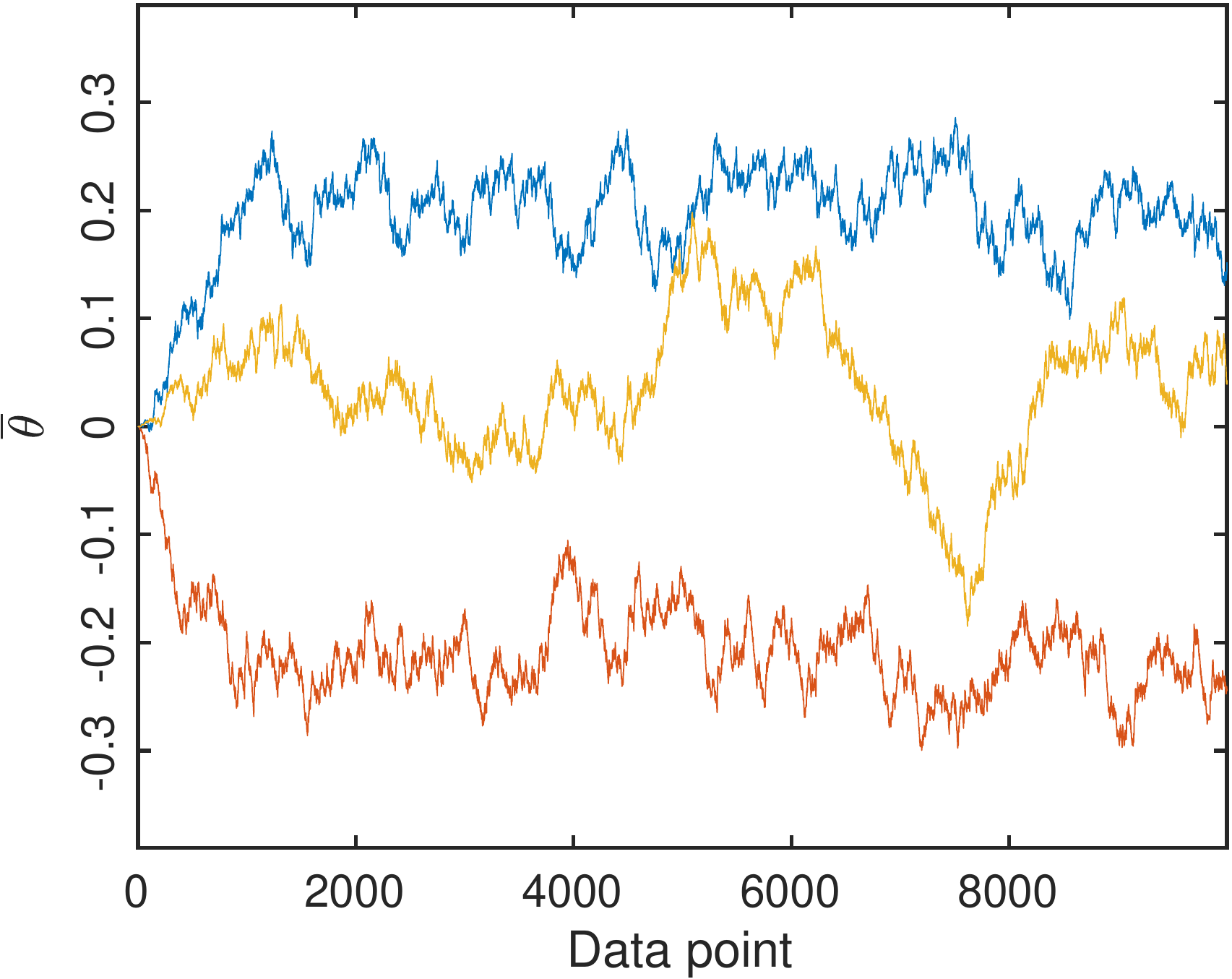}
    \includegraphics[width=0.49\textwidth]{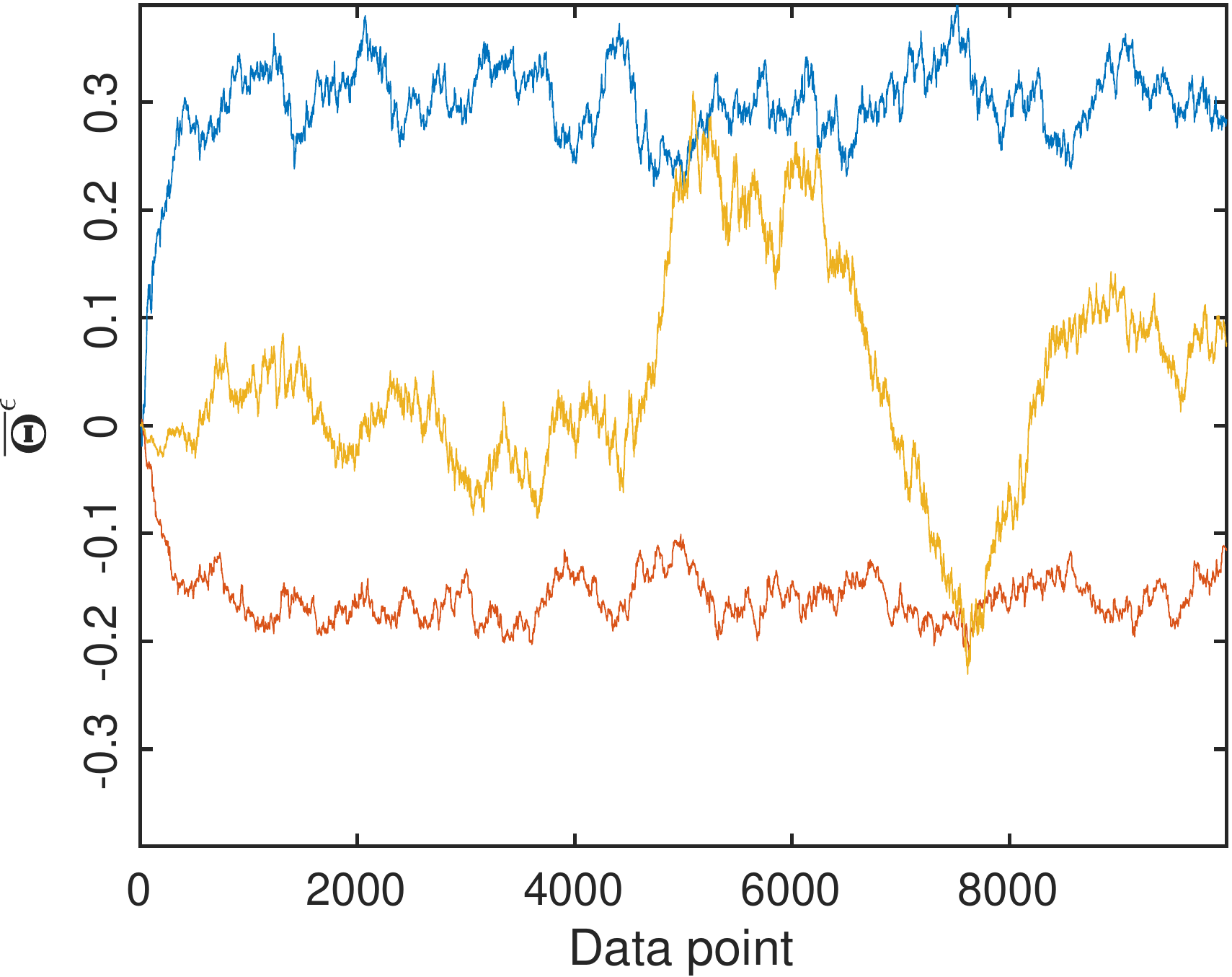}
    \includegraphics[width=0.49\textwidth]{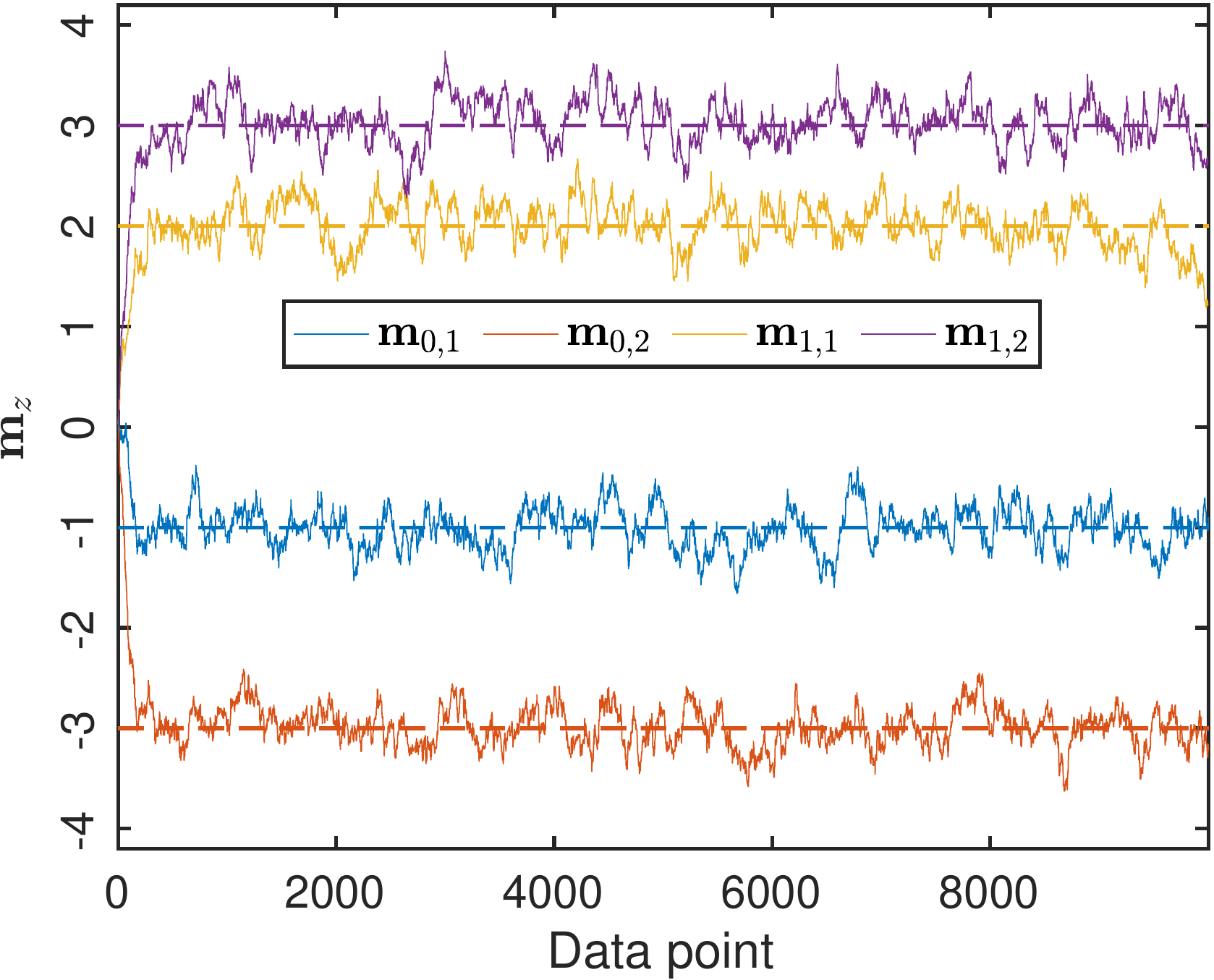}
    \includegraphics[width=0.49\textwidth]{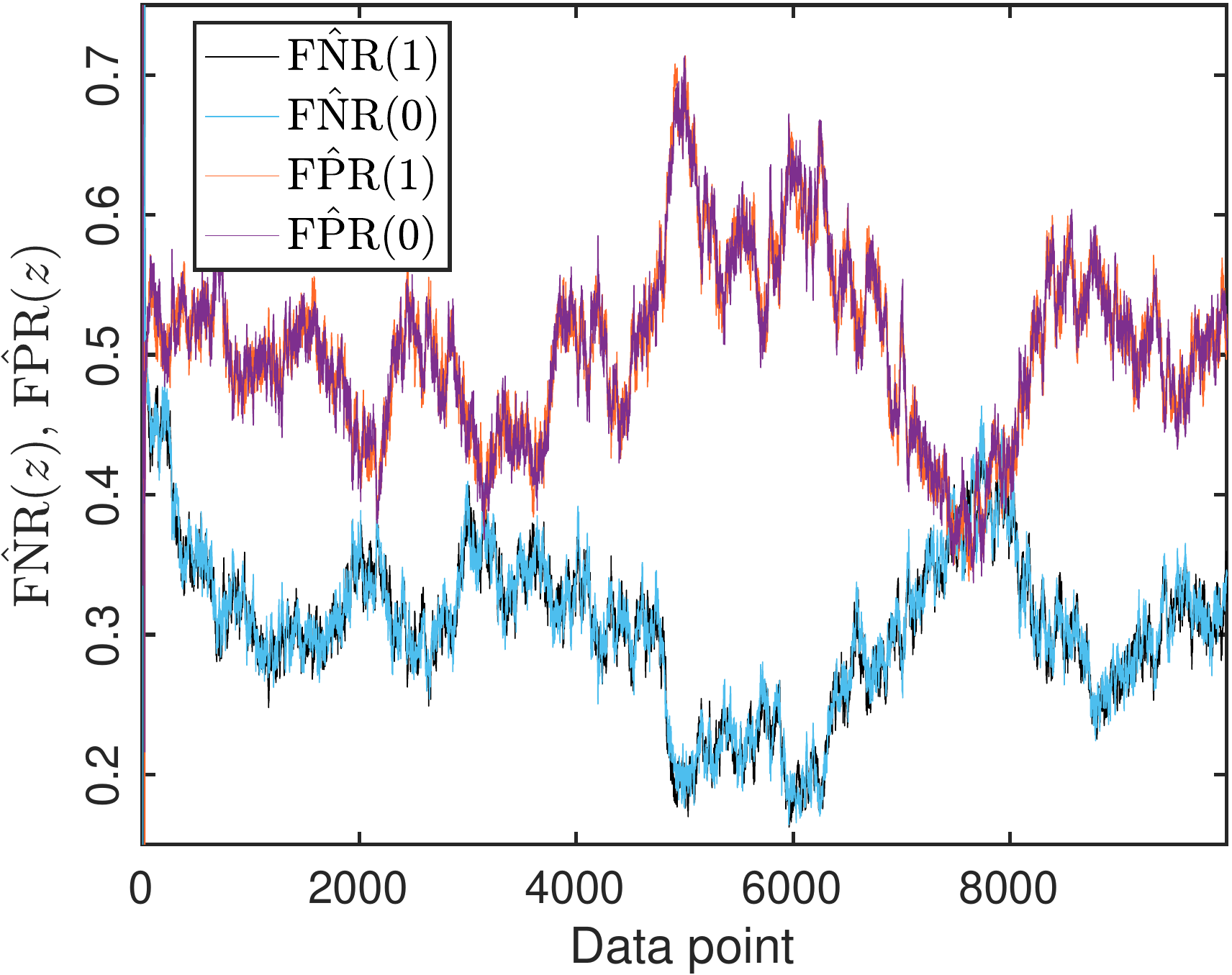}
    \caption{Plots of the tracked values of 
$\overline{\mathbf{\theta}}$ (top left),
$\overline{\mathbf{\Theta}}^\epsilon$ (top right), 
$\mathbf{m}_z$ (bottom left, shown with the true values $\mu_z$ as dashed lines),
and the estimated unbiased false prediction rates $\hat{\textrm{FNR}}(z)$ and $\hat{\textrm{FPR}}(z)$ (bottom right) for static parameters described in the text.}
    \label{fig:static}
\end{figure}

We use a low dimensional case of $N=3$.  The data is generated such that the static mean of $\mathcal{D}_0$ is $\mu_0=[-1;-3]$ while for $\mathcal{D}_1$ we have $\mu_1=[2;3]$; both use $\Sigma=[5,1;1,5]$.  To generate the simulated classifications $y_i$, each individual's $\mathbf{x}_i$ is generated via the appropriate $\mathcal{D}$, then the dot product of this feature vector with the vector $\mathbf{v}=[1,-1]$ is taken, such that $q_i=\mathbf{v}\mathbf{x}_i$.  Then, if $q_i>0$, we let $p_i=0.7$, while if $q_i<0$, we let $p_i=0.3$.  Finally, the true classification $y_i$ is Bernoulli with probability $p_i$.  Note, then, that the true classifications are not generated via a logistic function, though the feature vectors are in fact generated via multivariate normal distributions.  We generate $10,000$ such data points.  Initial values for the various tracking variable priors are $\overline{\theta}_{1|0}=[0;0;0]$, $C_{1|0}=0.0001I_3$, $\mathbf{m}_{z,1|0}=[0;0]$, $\lambda_{z,1|0}=\beta$, $\Phi_{z,1|0}=I_2$, and $\nu_{z,1|0}=4$.  We use $\epsilon=0.05$, $\alpha=0.85$, $Q=0.00001I_3$, and $\beta=49$.
Plots in Fig. \ref{fig:static} show the evolution of the estimated values of
$\overline{\mathbf{\theta}}$,
$\overline{\mathbf{\Theta}}$, 
$\mathbf{m}_z$,
and the false prediction rates $\hat{\textrm{FNR}}(z)$ and $\hat{\textrm{FPR}}(z)$.
The estimated covariance values are $\Sigma_0=[5.00,0.98;0.98,5.05]$ and $\Sigma_1=[5.11,1.05;1.05,5.02]$ after all points have been tracked.

Importantly, one can also determine the actual false prediction rates after the fact, obtaining false positive rates 0.48 and 0.48 and false negative rates 0.28 and 0.28 for sensitive values $z=0$ and $z=1$, respectively, over the final 9000 data points (we only use these points to allow for some stabilization of the algorithm before evaluating).  These are clearly within the tolerance requested, and match quite well to the estimated false prediction rates averaged over the last 9000 data points, which give 0.50 and 0.50 for false positive rates and 0.30 and 0.30 for false negative rates.  These essentially unbiased false prediction rates should be compared to the false prediction rates one would obtain if no bias elimination were employed (using a value of $\epsilon=2$), which are 0.62 and 0.23 for false positive rates and 0.11 and 0.46 for false negative rates.  The tradeoff of decreased bias is also a decrease in accuracy, however.  With no bias elimination, the accuracies are 0.69 and 0.67 for sensitive variable values 0 and 1, respectively, while our reduced-bias accuracies are 0.64 and 0.60.

The mean value of $\overline{\mathbf{\theta}}$ over the last 9000 points is $[0.21,-0.21,0.04]$, while the average value of $\overline{\mathbf{\Theta}}^\epsilon$ is $[0.30,-0.16,0.05]$.  
A standard logistic regression over this set of data yields a coefficient vector $[0.21, -0.22,0.04]$ almost identical to the mean of $\overline{\mathbf{\theta}}$ that we obtain, showing that our logistic tracker works as anticipated.  As a point of comparison to other bias-reducing logistic classification methods, we also apply the algorithm of \cite{zafar2017fairness} to this dataset, using the form that attempts to minimize differences in both false positive and false negative rates simultaneously, with a covariance threshold multiplicative factor (see \cite{zafar2017fairness} for definition) of $m=0.0005$.  The results in that case are a coefficient vector of $[0.23,-0.13,0.09]$, leading to false positive rates of 0.53 and 0.51, and false negative rates of 0.22 and 0.27.  While certainly less biased than the standard logistic classifier, the difference in false prediction rates with the algorithm of \cite{zafar2017fairness} is notably larger than in our case.  This is likely due to the fact that our algorithm is directly attempting to equalize the false prediction rates, while \cite{zafar2017fairness} uses a proxy measure to achieve the same goal.  Of course, our algorithm is much slower than that of \cite{zafar2017fairness} due to the several Monte Carlo steps involved.


\subsection{Dynamic parameters} \label{sec:symswap}
The scenario here is very similar to that used in the static parameter case above.  The exception is that, after the first 1000 data points, the means $\mu_0$ and $\mu_1$ begin to linearly drift with each new data point, such that the two values have exactly swapped by the 10,000th data point.  Specifically, we use $\mu_{0,i}=[-1;-3]$ and $\mu_{1,i}=[2;3]$ for $i\leq 1000$ while $\mu_{0,i}=[-1;-3]+[3;6](i-1000)/9000$ and $\mu_{1,i}=[2;3]-[3;6](i-1000)/9000$ for $i>1000$.  This highly contrived scenario allows us to focus on a situation in which, when considering the final 9000 data points all together, we expect to see little difference in false prediction rates between the two sensitive variable values even if a standard logistic regression is used.  However, it is clear that will be, in general, \emph{instantaneous} bias in the predictions.  This bias will switch between the two groups over the course of the observations, making the overall false prediction rates roughly equivalent.

\begin{figure}
    \centering
    \includegraphics[width=0.49\textwidth]{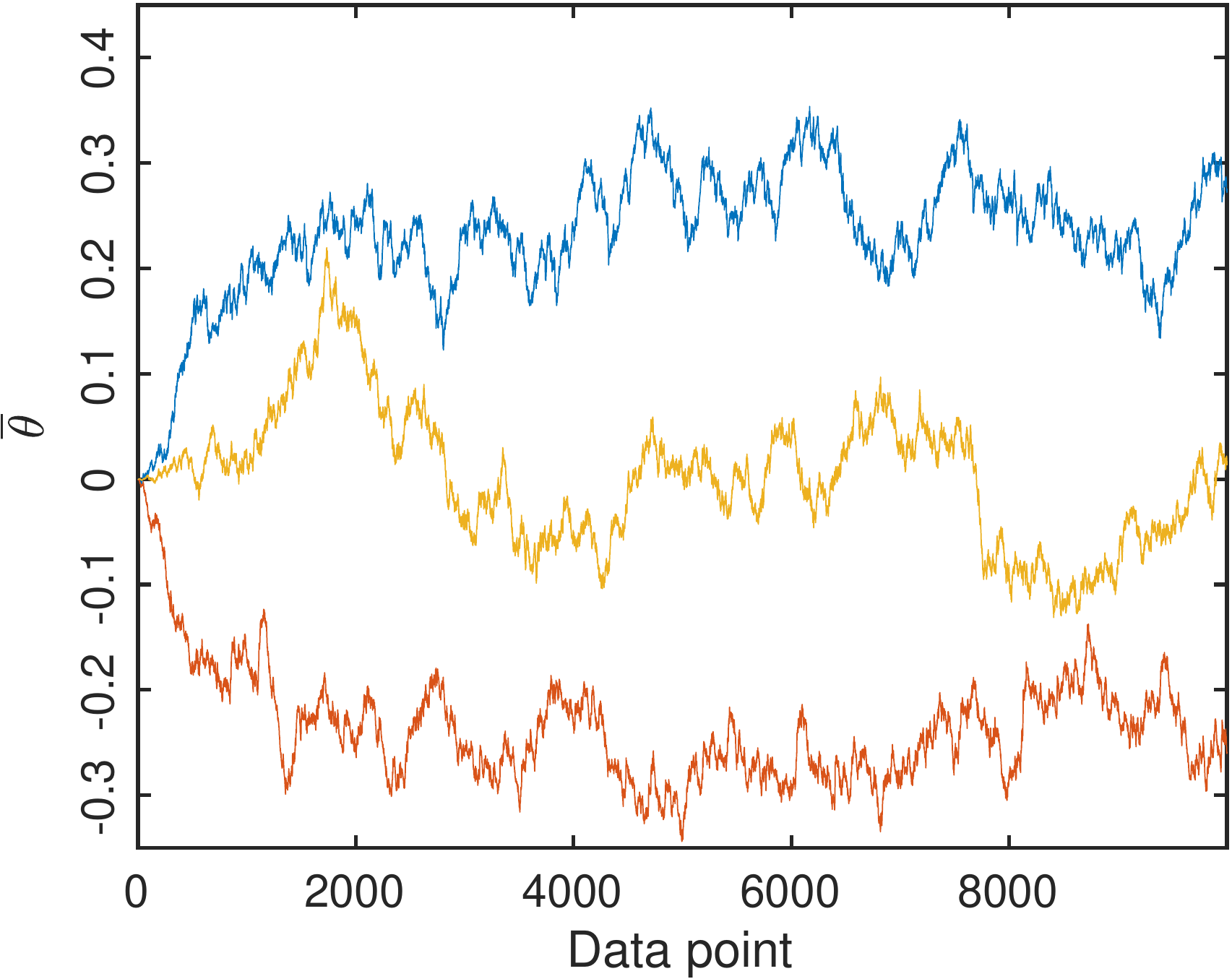}
    \includegraphics[width=0.49\textwidth]{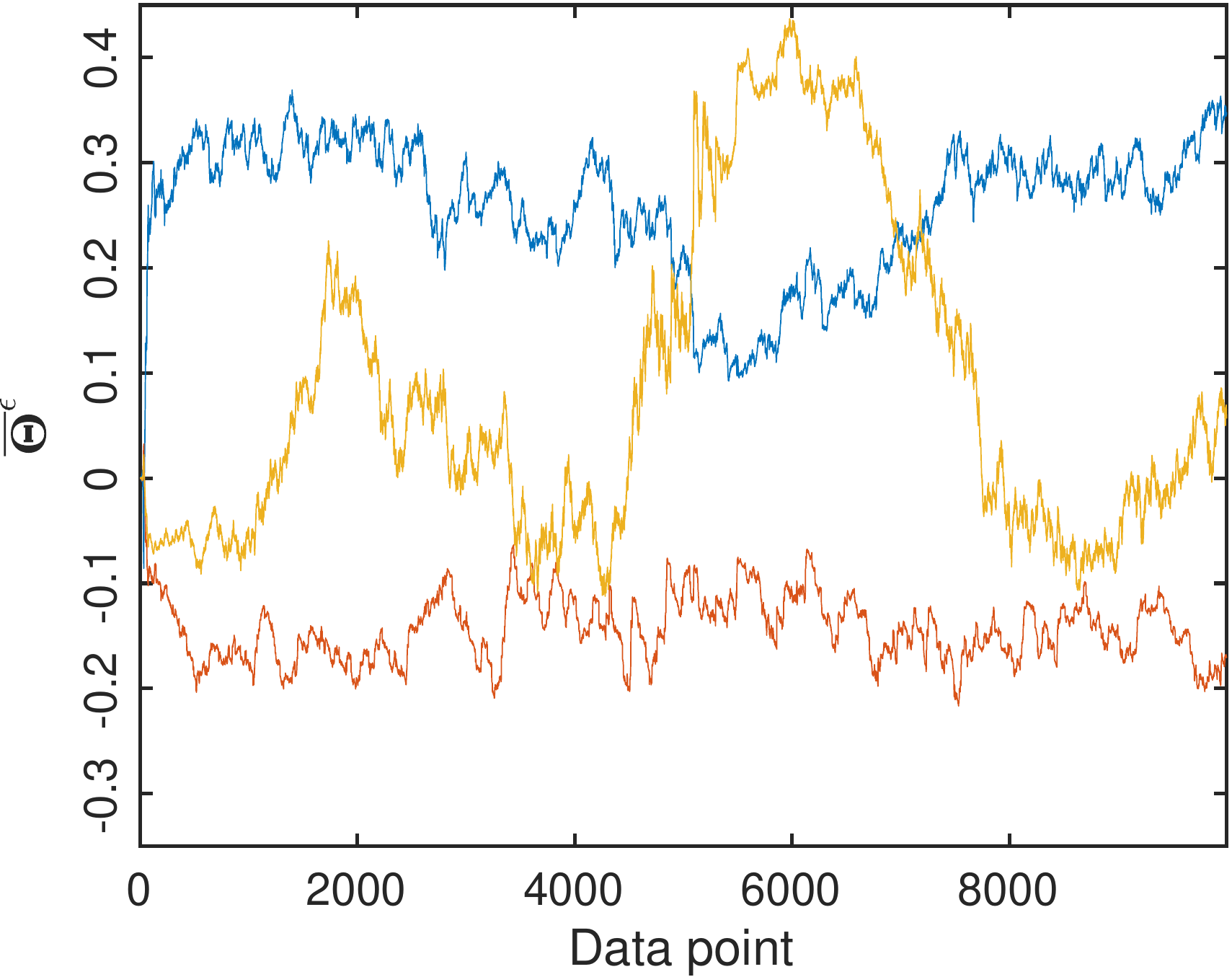}
    \includegraphics[width=0.49\textwidth]{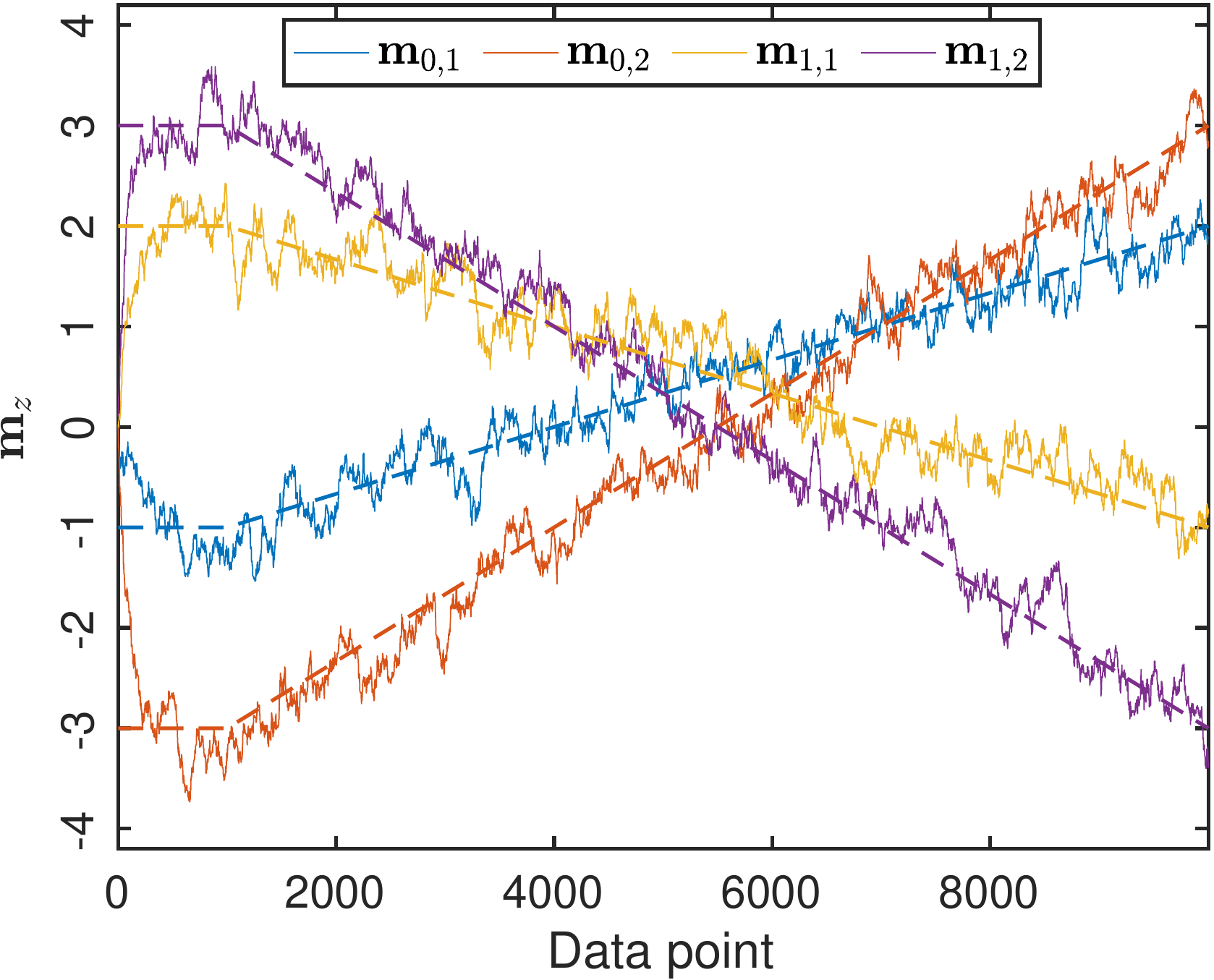}
    \includegraphics[width=0.49\textwidth]{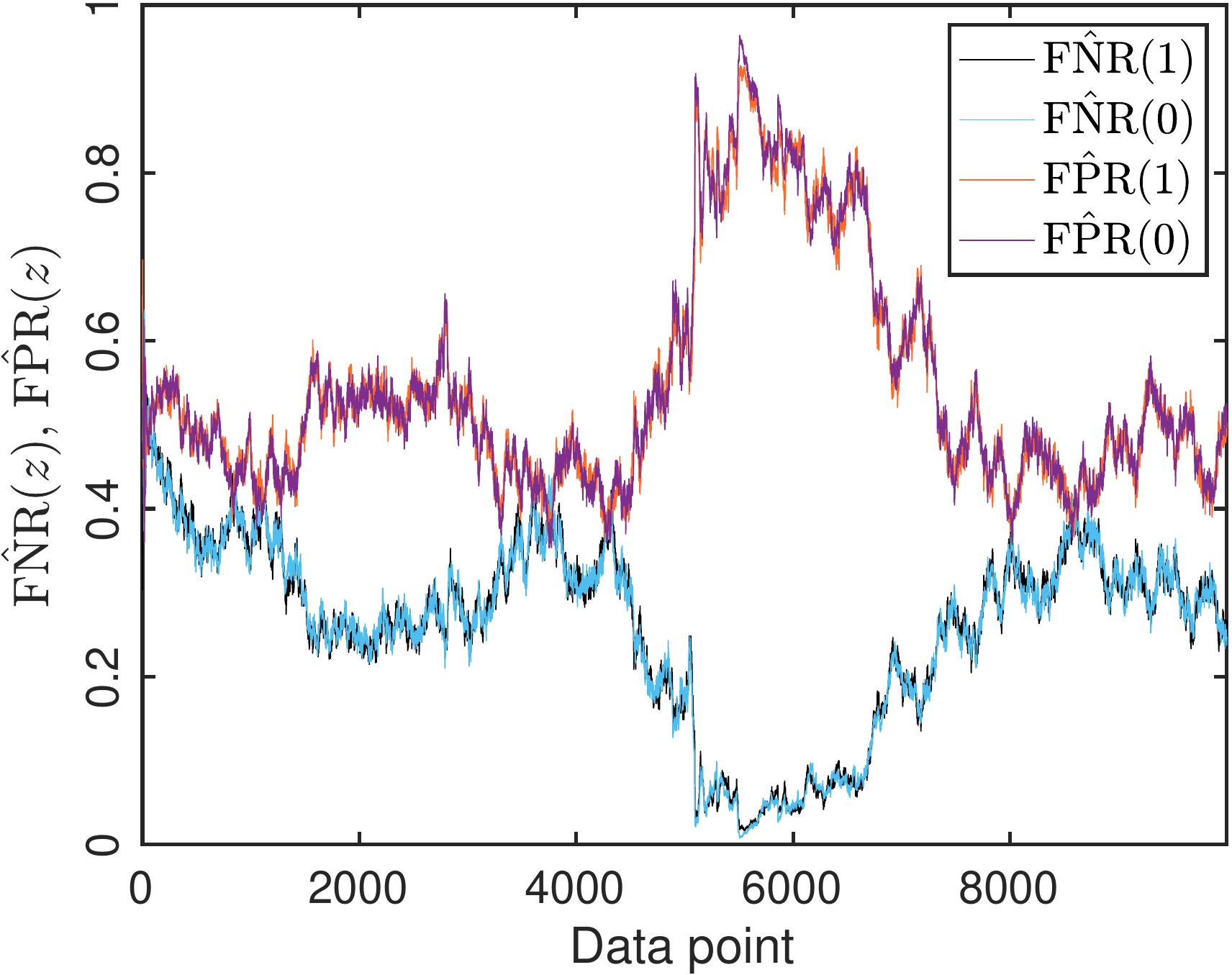}
    \caption{Plots of the tracked values of 
$\overline{\mathbf{\theta}}$ (top left),
$\overline{\mathbf{\Theta}}^\epsilon$ (top right), 
$\mathbf{m}_z$ (bottom left, shown with the true values $\mu_z$ as dashed lines),
and the estimated unbiased false prediction rates $\hat{\textrm{FNR}}$ and $\hat{\textrm{FPR}}$ (bottom right) for dynamic parameters described in the text.}
    \label{fig:dynamic2}
\end{figure}

We run our tracking algorithm with the same parameters and initial values as in the static case, and present results in Fig.~\ref{fig:dynamic2}.  Here, our reduced bias false prediction rates as measured over the final 9000 data points are 0.52 and 0.54 for false positives and 0.22 and 0.23 for false negatives for sensitive values $z=0$ and $z=1$, respectively; accuracies are 0.64 and 0.62.  However, as shown in Fig.~\ref{fig:dynamic2}, the estimated instantaneous false prediction rates vary significantly over the course of the tracking, albeit in such a way that the estimated bias is always within our tolerance $\epsilon$.

However, by design, when we run this scenario through our algorithm using the large value $\epsilon=2.0$, in which case no bias removal is actually attempted, the overall observed results still appear effectively unbiased.  Over the final 9000 observations, we obtain 0.38 and 0.37 for false positive rates and 0.24 and 0.27 for false negative rates for sensitive values $z=0$ and $z=1$, respectively; accuracies are 0.69 and 0.69.  The mean value of $\overline{\mathbf{\theta}}$ over the last 9000 points in this case is $[0.25,-0.25,0.00]$.  All of these values are very close to those obtained for a standard logistic regression over these same points.  Even the method of \cite{zafar2017fairness} again using a covariance threshold multiplicative factor of $m=0.0005$ yields 0.36 and 0.35 for false positive rates and 0.23 and 0.24 for false negative rates for sensitive values $z=0$ and $z=1$, respectively, with accuracies 0.71 and 0.71.  These are only slightly different than those of the standard logistic regression, since it contains almost no overall bias in the first place.

\begin{figure}
    \centering
    \includegraphics[width=0.49\textwidth]{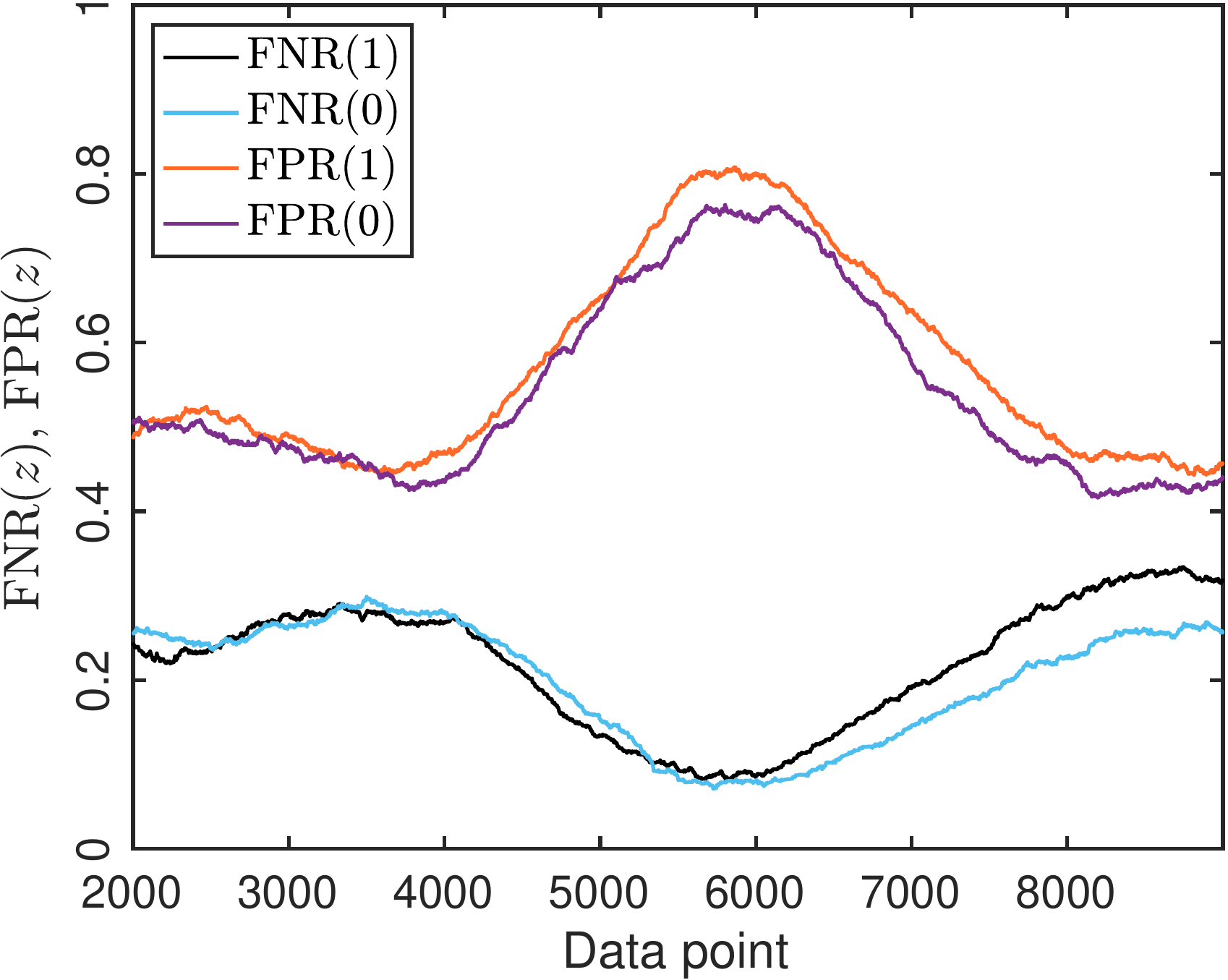}
    \includegraphics[width=0.49\textwidth]{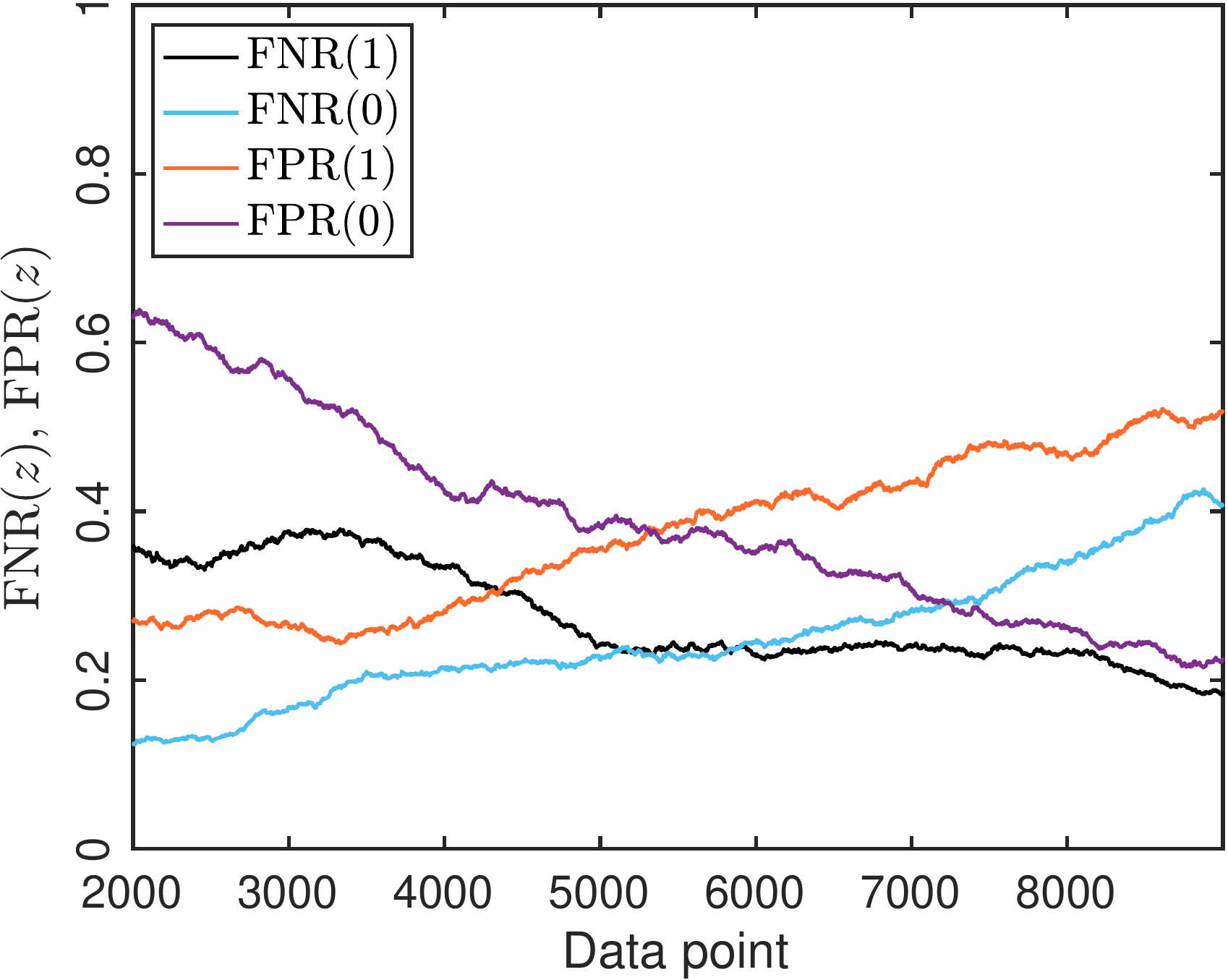}
    \caption{Plots of the observed false prediction rates computed via a moving-window average of width 2000 events for our dynamically unbiased predictions (left panel) and for effectively static logistic predictions (right panel).}
    \label{fig:dynamic2b}
\end{figure}

But, to illustrate the difference between these static methods and our dynamic method on this dynamic dataset, we plot in Fig.~\ref{fig:dynamic2b} the observed false prediction rates when calculated via a symmetric moving-window average of width 2000 events.  In the case of our dynamic method with $\epsilon=0.05$ (left panel), these moving averages show relatively small bias between the two protected variable values, and are similar to the estimated false prediction rates shown in Fig.~\ref{fig:dynamic2}.  However, when the same moving-window average is applied to predictions made by our tracking algorithm but with $\epsilon=2.0$ (right panel), which mimics a static logistic regression that appears unbiased when considering the entire dataset at once as discussed above, we clearly see large difference in false prediction rates between the two protected variable groups.




\section{Results on ProPublica COMPAS dataset}
To illustrate the ability of our algorithm to analyze real-world datasets, we applied it to the often-used ProPublica COMPAS 2-year recidivism dataset \cite{propublica}.  In analyzing the dataset, we have chosen race as the sensitive variable, and only analyzed that subset of the data in which the race is listed as either ``African-American'' or ``Caucasian''.  After selecting this subset, and removing a few points in the same way as described in \cite{propublica}, we are left with 5278 entries.

For our features, we have analyzed two scenarios.  In both scenarios, we use ``sex'' (categorical, male=0 or female=1), ``age\_cat'' (using two categorical variables ``Less than 25'' and ``Greater than 45''), ``priors\_count'' (number of prior crimes), and ``c\_charge\_degree'' (categorical, felony=0 or misdemeanor=1).  In the first scenario, these are the only features considered, while in the second scenario we additional directly consider ``race'' as the final feature (categorical, Caucasian=0 or African-American=1); this is done to match what some other have considered when analyzing this particular dataset.  Importantly, we also analyze this dataset after sorting the entries by ``compas\_screening\_date'' from earliest to latest.  This is done because our algorithm is specifically developed to allow for temporally evolving scenarios, so we have evaluated it as such.  All initial parameters of the algorithm are the same as used in the synthetic data above, with the exception that we use $\alpha=0.65$ here, since (as we will see), some of the accuracies are quite low.

Table \ref{tab:compas_full} lists the results of our algorithm, as well as the those of \cite{zafar2017fairness} for both sets of features (with and without race), and both with and without any bias constraints applied; for the algorithm of \cite{zafar2017fairness}, the bias constraint was on both false positive and false negative rates simultaneously, using a covariance threshold multiplicative factor of $m= 0.000001$.  The results here are computed over the second half of the dataset only; in the case of the Zafar algorithm, only the first half of the dataset is used for training.  The results clearly show that, without any constraints, there is bias between false prediction rates of Caucasians vs African-Americans: false negative rates for Caucasians are higher than those for African-Americans, and the opposite is true of false positive rates.  When constraints are added, both algorithms greatly reduce these differences, with ours accomplishing our goal of $\epsilon<0.05$ in all cases, but the Zafar algorithm is unable to reduce the bias levels to within this same tolerance.  This is not necessarily surprising, as the Zafar algorithm only uses a proxy measure for bias, rather than an explicit calculation of the expected level as our own algorithm employs.  Within the table, we have specifically noted the number of positive predictions given in each case.  This is displayed to illustrate the fact that, in general, the constrained algorithms accomplish their goal by classifying many more individuals as negative than in the unconstrained case.  The most extreme example of this is in the Zafar algorithm with race not included as a feature, in which case only 115, or 4.4\%, of the 2639 predictions made are positive.  This leads to an overall accuracy of only 0.48.  It seems that, if not explicitly using race to make predictions, a static unbiased classifier like Zafar can only really treat this data by classifying essentially everyone in the same way (negative).

\begin{landscape}
\begin{table}[]
    \centering
    \begin{tabular}{c|c|c|c|c|c|c|c|c|c}
         Algorithm & Features & Pred. Pos. & Overall Acc. & C. Acc. & C. FNR & C. FPR & A.A. Acc. & A.A. FNR & A.A. FPR  \\
         Ours, uncons. & w/ race & 1023 & 0.75 & 0.73 & 0.50 & 0.05 & 0.76 & 0.31 & 0.13 \\
         Zafar, uncons. & w/ race & 668 & 0.61 & 0.59 & 0.82 & 0.03 & 0.63 & 0.51 & 0.15 \\
         Ours, cons. & w/ race & 718 & 0.68 & 0.71 & 0.54 & 0.05 & 0.66 & 0.54 & 0.03 \\
         Zafar, cons. & w/ race & 754 & 0.61 & 0.61 & 0.63 & 0.18 & 0.60 & 0.58 & 0.11 \\
         Ours, uncons. & w/o race & 1000 & 0.74 & 0.73 & 0.49 & 0.06 & 0.75 & 0.34 & 0.12 \\
         Zafar, uncons. & w/o race & 607 & 0.60 & 0.60 & 0.78 & 0.06 & 0.60 & 0.59 & 0.12 \\
         Ours, cons. & w/o race & 655 & 0.69 & 0.72 & 0.57 & 0.00 & 0.67 & 0.54 & 0.00 \\
         Zafar, cons. & w/o race & 115 & 0.48 & 0.53 & 0.97 & 0.00 & 0.45 & 0.91 & 0.01 \\
    \end{tabular}
    \caption{Results of our algorithm and those of Zafar et al. \cite{zafar2017fairness} on the full subset of the ProPublica dataset described in the text, under various different combinations of feature vectors (both with and without race) and constraints.  Here, ``C.'' is ``Caucasian'' and ``A.A.'' is ``African-American''.} 
    \label{tab:compas_full}
\end{table}
\end{landscape}

\begin{figure}
    \centering
    \includegraphics[width=0.49\textwidth]{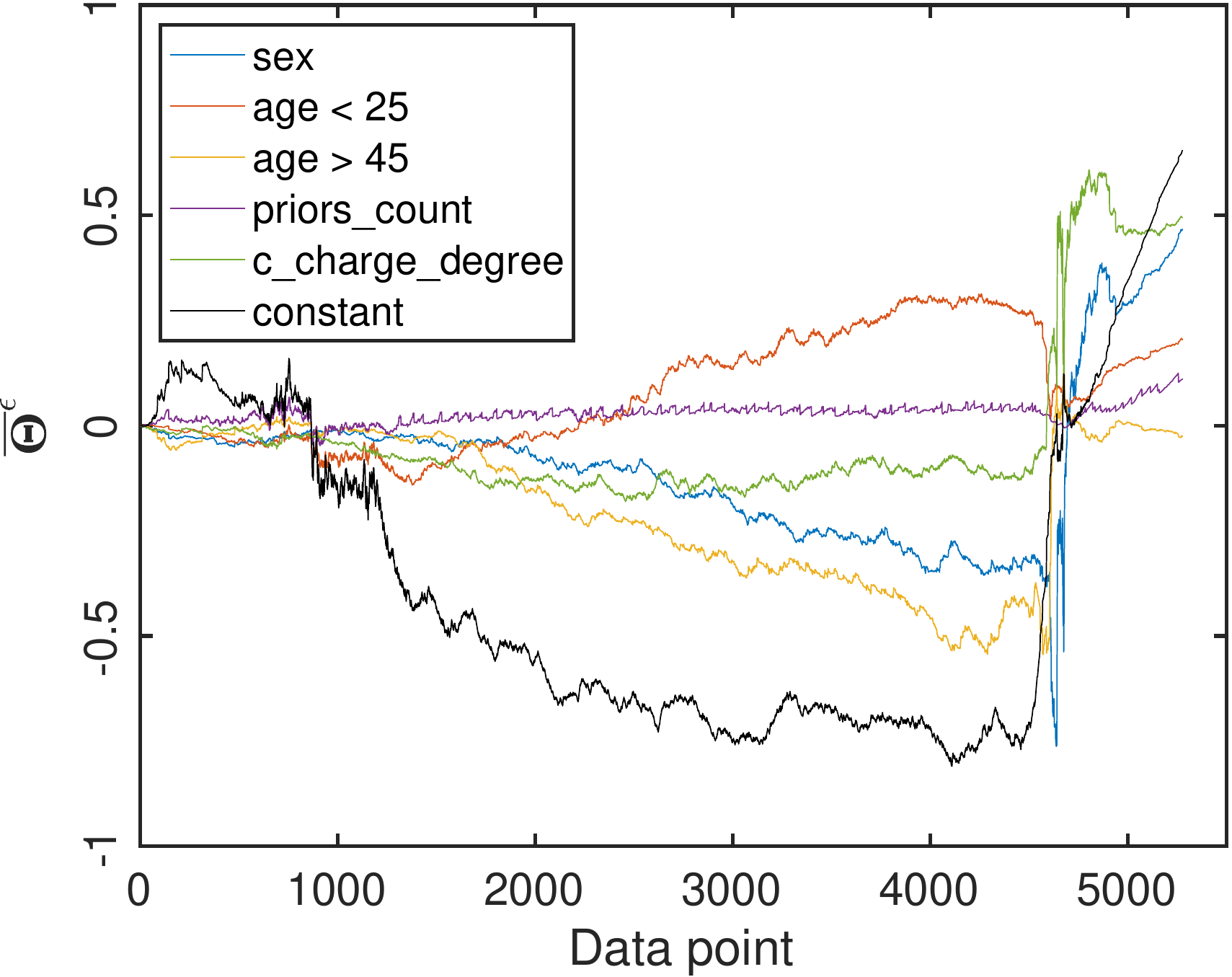}
    \includegraphics[width=0.49\textwidth]{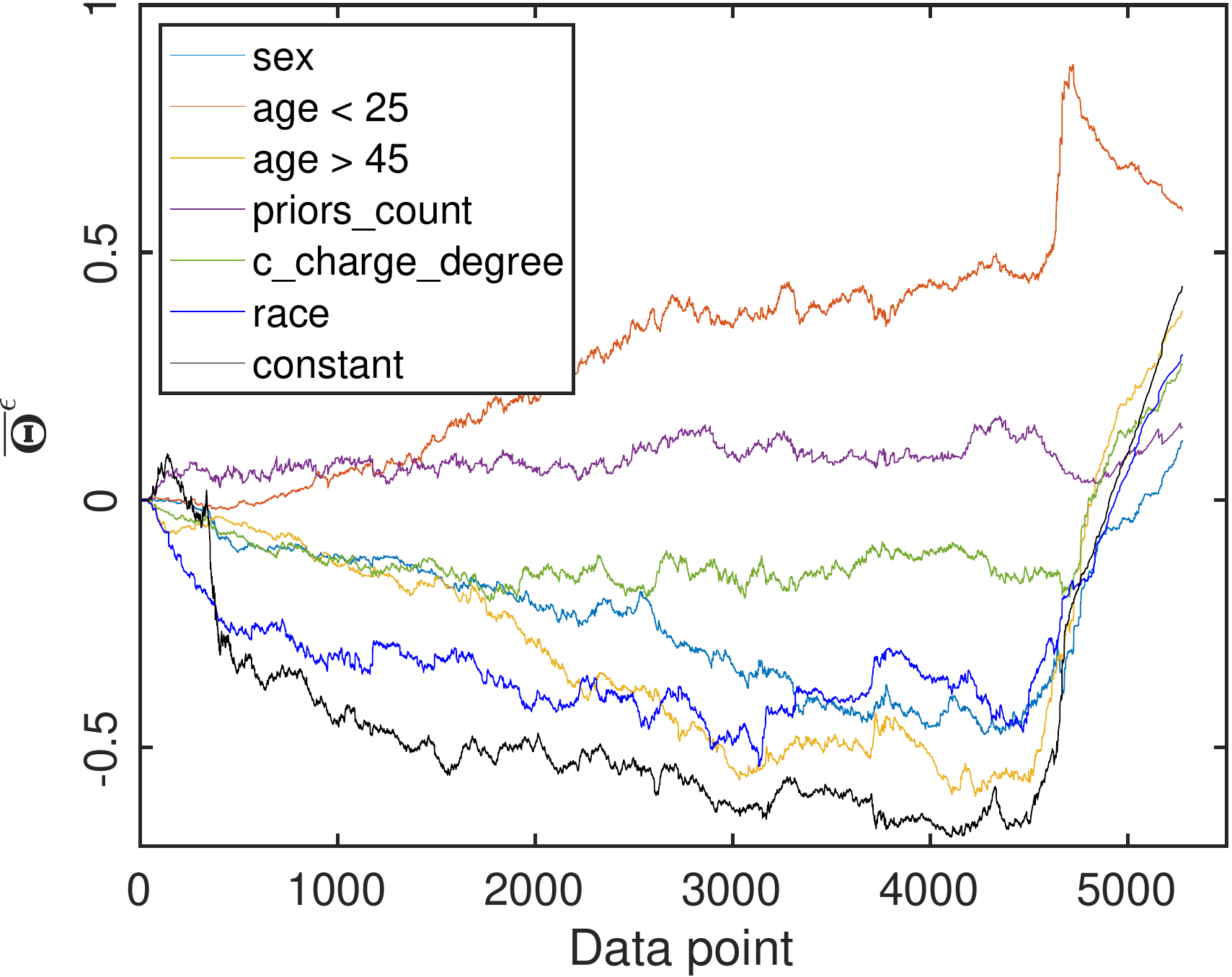}
    \caption{Plots of the tracked values of 
$\overline{\mathbf{\Theta}}^\epsilon$ for the full ProPublica COMPAS dataset both without (left panel) and with (right panel) race as a feature.}
    \label{fig:compas_full}
\end{figure}

Interestingly, though, our own constrained classifier in the no-race case still makes 655 positive predictions, and ends up with an overall accuracy of 0.69, not much lower than the accuracy of 0.74 that we achieve with the same features in the unconstrained case.  Somewhat amazingly, our algorithm does this with a false positive rate of 0 for both races - all 655 positive predictions were correct.  Also, our algorithm displays significantly better overall accuracy than Zafar across the board here.  To help understand how our algorithm achieves such results, we plot in Fig.~\ref{fig:compas_full} the evolving estimates of the unbiased coefficients $\overline{\mathbf{\Theta}}$ that our algorithm produces in both the with-race and without-race scenarios.  As can be clearly seen, there is an abrupt and very large change in estimated coefficients at around data point number 4600 in each case.  Upon investigating the data directly, it was observed that every datapoint starting at number 4512 (with ``compas\_screening\_date'' of April 2, 2014) is classified as a recidivist.  This data, if it is to be believed, is then a perfect example of a scenario in which a temporal trend is important to the classification task, which our algorithm handles quite naturally.  However, it is likely that these datapoints at the end of the data set are in fact erroneously classified; in fact, this precise observation has been pointed out elsewhere \cite{compas_bad}.    

In light of this observation, we have run all of our analyses on a further subset of the COMPAS data, removing all those entries with a `compas\_screening\_date'' on or after April 2, 2014, as suggested by \cite{compas_bad}.  This leaves 4511 data points in this second analysis.  The results are shown in Table \ref{tab:compas_trim}.  In comparison to Table \ref{tab:compas_full}, the results in Table \ref{tab:compas_trim} show that the overall accuracy of our algorithm drops, while those of Zafar increase, as one would expect.  Comparing our algorithm to that of Zafar, we find that with the exception of the constrained, with-race case, results are roughly similar, though Zafar generally has slightly higher accuracy at the expense of somewhat greater bias.  The constrained, with-race results are quite different between our two algorithms, however.  In this case, the overall accuracy of the two algorithms are on par, but these are acheived in very different ways.  Our algorithm classifies far fewer individuals as positive than Zafar in this case, ending up with quite high false negative rates and very low false positive rates, with little difference between the two races.  On the other hand, the Zafar algorithm actually classifies more individuals as positive in this case than the unconstrained, with-race case, yielding a moderately high false negative rate as well as a low but notable false positive rate, and with still significant disparity between the races, albeit with both false prediction rates for Caucasians now greater than the corresponding rates for African-Americans.

\begin{landscape}
\begin{table}[]
    \centering
    \begin{tabular}{c|c|c|c|c|c|c|c|c|c}
         Algorithm & Features & Pred. Pos. & Overall Acc. & C. Acc. & C. FNR & C. FPR & A.A. Acc. & A.A. FNR & A.A. FPR  \\
         Ours, uncons. & w/ race & 408 & 0.70 & 0.72 & 0.83 & 0.05 & 0.68 & 0.58 & 0.13 \\
         Zafar, uncons. & w/ race & 471 & 0.70 & 0.71 & 0.85 & 0.05 & 0.70 & 0.50 & 0.16 \\
         Ours, cons. & w/ race & 197 & 0.67 & 0.72 & 0.84 & 0.05 & 0.63 & 0.83 & 0.03 \\
         Zafar, cons. & w/ race & 545 & 0.69 & 0.68 & 0.64 & 0.18 & 0.69 & 0.57 & 0.11 \\
         Ours, uncons. & w/o race & 395 & 0.69 & 0.72 & 0.82 & 0.06 & 0.68 & 0.61 & 0.11 \\
         Zafar, uncons. & w/o race & 416 & 0.70 & 0.71 & 0.83 & 0.06 & 0.69 & 0.57 & 0.12 \\
         Ours, cons. & w/o race & 17 & 0.63 & 0.70 & 1.00 & 0.00 & 0.59 & 0.97 & 0.00 \\
         Zafar, cons. & w/o race & 79 & 0.65 & 0.71 & 0.96 & 0.00 & 0.61 & 0.90 & 0.01 \\
    \end{tabular}
    \caption{Results of our algorithm and those of Zafar et al. \cite{zafar2017fairness} on the restricted subset of the ProPublica dataset described in the text with all data points with ``compas\_screening\_date'' of April 2, 2014 or later removed, under various different combinations of feature vectors (both with and without race) and constraints.  Here, ``C.'' is ``Caucasian'' and ``A.A.'' is ``African-American''.} 
    \label{tab:compas_trim}
\end{table}
\end{landscape}

\section{Conclusions}
In this work we introduced a fully Bayesian tracking algorithm for fairness-aware logistic regression.  The model sequentially tracks potential changes in the distribution of features, along with false positive and negative rates, and dynamically adjusts the model to mitigate disparate misclassification.  We demonstrated the effectiveness of the algorithm on synthetic and recidivism datasets, showing improved performance with regard to disparate misclassification compared to bias reducing methods that are trained in batch offline.   

The present methodology has several limitations that should be noted.  Here we assumed that class labels were fully observed in real time, whereas in practice some labels are unobserved (for example recidivism may go unobserved) and other labels may only be available after some delay.  The method also introduces potential bias as it relates to individual fairness \cite{dwork2012fairness}, the notion that individuals with similar features should receive similar algorithmic scores and decisions.  Due to the dynamic nature of the present algorithm, an individual at an earlier time may receive a different decision than an individual with similar features at a later time.

One direction for future research would be to extend the present methodology to other situations where algorithmic decisions are made sequentially under changes in the underlying distribution of the data.  For example, issues of bias and fairness may arise in other criminal justice applications beyond parole and bail decisions, including traffic stops and hotspot policing based on spatial crime forecasts.  These scenarios pose additional challenges, where existing models include decision thresholds that must be inferred from data \cite{simoiu2017problem} and spatial-temporal dynamics with potential feedback \cite{lum2016predict,brantingham2018does,mohler2018penalized}. 

\section{Acknowledgements}
This research was supported by NSF grants SCC-1737585, ATD-1737996 and ATD-1737925.

\bibliographystyle{plain}
\bibliography{biblio}
\end{document}